\documentclass[12pt,a4paper,english]{article}
\usepackage[T1]{fontenc}
\usepackage[latin2]{inputenc}
\usepackage[english]{babel}
\usepackage{amsmath}
\usepackage{amsfonts}
\usepackage{indentfirst}
\usepackage[dvips]{graphicx}
\usepackage{youngtab}

\newcommand{\bea}{\begin{eqnarray}}
\newcommand{\eea}{\end{eqnarray}}
\newcommand{\be}{\begin{equation}}
\newcommand{\ee}{\end{equation}}

\newcommand{\hf}{\frac{1}{2}}

\def\Tr{{\rm Tr}}

\def\G{\Gamma}

\def\hs{h(\square)}

\newcommand{\cF}{{\cal F }}

\newcommand{\cN}{{\cal N }}

\newcommand{\Li}{{\rm Li}}

\begin{document}

\sloppy


\begin{flushright}
\begin{tabular}{l}
BONN-TH-2009-03 \\

\\ [.3in]
\end{tabular}
\end{flushright}

\begin{center}
\Large{ \bf Matrix models for $2^*$ theories}
\end{center}

\begin{center}

\bigskip

Piotr Su{\l}kowski

\bigskip

\medskip

\emph{Physikalisches Institut der Universit{\"a}t Bonn and Bethe Center for Theoretical Physics,} \\
\emph{Nussallee 12, 53115 Bonn, Germany} \\  [2mm]
\emph{and} \\ [2mm]
\emph{So{\l}tan Institute for Nuclear Studies, ul. Ho\.za 69, 00-681 Warsaw, Poland} \\ 

\bigskip

\emph{Piotr.Sulkowski@fuw.edu.pl}

\bigskip

\smallskip
 \vskip .4in \centerline{\bf Abstract}
\smallskip

\end{center}

We show how to represent a class of expressions involving discrete sums over partitions as matrix models. We apply this technique to the partition functions of $2^*$ theories, i.e. Seiberg-Witten theories with the massive hypermultiplet in the adjoint representation. We consider theories in four, five and six dimensions, and obtain new matrix models respectively of rational, trigonometric and elliptic type. The matrix models for five- and six-dimensional $U(1)$ theories are derived from the topological vertex construction related to curves of genus one and two.


\newpage

\section{Introduction}

Matrix models have become an important ingredient of theoretical physics over the last two decades. The information about the matrix model is encoded in the partition function 
\be
Z = \int \mathcal{D}M\, e^{-\frac{1}{\hbar} \Tr \, V(M)}, \label{mmodel} 
\ee
where the measure $\mathcal{D}M$ represents an integral over an appropriate set of $N\times N$ matrices $M$. The solution of the matrix model is encoded in the spectral curve which arises from eigenvalues condensing in the extrema of the potential $V$ in the large $N$ limit. Taking at the same time $\hbar\to 0$ with the 't Hooft coupling $g_s N = const$, the solution of the matrix integral is given by coefficients $F_g$ which arise in perturbative expansion in $\hbar$ of $F=\log Z$.

In this paper we show how to represent a class of expressions involving sums over two-dimensional partitions as matrix models. Such expressions often arise as partition functions of $\cN=2$ supersymmetric gauge theories and topological strings. In particular we will apply this technique to so-called $2^*$ theories, i.e. Seiberg-Witten theories with the massive hypermultiplet in the adjoint representation.

To start with we recall some other relations established between matrix models and supersymmetric field theories, and explain how our approach differs from them. First of all, one such relation is the Dijkgraaf-Vafa correspondence \cite{DV}. According to this correspondence the low-energy effective superpotential of $\mathcal{N}=1$ theory is computed by the matrix model, whose potential is identified with the tree level superpotential. While immense amount of work has been done to verify and apply this correspondence in the $\mathcal{N}=1$ context, it is also possible to extract from it some information about $\mathcal{N}=2$ theories. To do that one has to consider $\mathcal{N}=2$ theory broken to $\mathcal{N}=1$ by a superpotential treated as a perturbation, adjust the superpotential in a way which allows to choose a relevant point in the Coulomb branch, apply the Dijkgraaf-Vafa conjecture for the $\mathcal{N}=1$ situation and eventually identify certain parameters and set the deformation to zero to get $\mathcal{N}=2$ data \cite{cachazo-vafa}. In this way the prepotential $\mathcal{F}_0$ of $\mathcal{N}=2$ theories and its gravitational correction $\mathcal{F}_1$ have been obtained from matrix model calculations in some cases \cite{gaugedmm,kmt,DST}. 

The results of Dijkgraaf of Vafa allow us to get the information related to F-terms in $\cN=1$ theories due to underlying holomorphicity. Another sector of these theories is encoded in D-terms whose full determination would require going beyond the holomorphic regime. Nonetheless just the holomorphic information is sufficient to solve more rigid, but still non-trivial $\cN=2$ theories \cite{SW-1,SW-2}. Seiberg-Witten solution of these theories is encoded in a family of Seiberg-Witten curves. The research in this direction culminated in the derivation of the Nekrasov partition function \cite{Nek,Nek-Ok} for $\cN=2$ theories. These partition functions $Z_{Nekrasov}$ arise from the localization and are given by sums over two-dimensional partitions which label instanton configurations. They can be expanded as
$$
Z_{Nekrasov} = e^{\sum_{g=0}^{\infty} \cF_g \hbar^{2g-2}}
$$
and apart from the prepotential $\cF_0$ they also encode the whole series of gravitational corrections $\cF_g$. 

Gauge theories with $\cN=2$ supersymmetry are also intimately related to string theory on local Calabi-Yau manifolds via geometric engineering \cite{geom-eng}. In particular partition functions of the large radius A-model topological strings on toric Calabi-Yau manifolds agree with Nekrasov partition functions for 5-dimensional gauge theories upon appropriate identification of parameters, and reduce to 4-dimensional result in a proper limit \cite{suN-Amir,HIV,Zhou2003,Zhou2004}. As the problem of computing the large radius topological string partition functions on toric manifolds is essentially solved by the topological vertex theory \cite{vertex,adkmv}, so are the corresponding gauge theories which can be geometrically engineered. In this correspondence the Seiberg-Witten curve is encoded in the target space of the mirror B-model geometry. More generally, each toric threefold encoded in the diagram constructed from topological vertices is dual to a geometry based on a corresponding Riemann surface.

One immediately notices similarities between matrix models, $\cN=2$ theories, and topological strings. In each system there is a Riemann surface which governs the solution: the spectral curve, the Seiberg-Witten, and the curve in the B-model target space. It is advantageous to relate explicitly gauge theories and topological strings to matrix models, both conceptually, as well as with a hope to apply techniques from one side of such an equivalence to the other side. The Dijkgraaf-Vafa correspondence can be regarded as one such possibility. However this allows to obtain only $\cF_0$ and $\cF_1$ data from the corresponding matrix model $F_0$ and $F_1$, while for $g\geq 2$ gauge theory $\cF_{g}$'s do not agree with matrix model $F_g$'s \cite{kmt,kmr}. Moreover this lowest genus $\cN=2$ answer is not explicitly given by the matrix integral in this case, but it arises only after certain identifications of parameters. We also note that the idea of c-deformation was advanced \cite{c-deform} to relate the all genus partition function of the matrix model, including non-planar diagrams, to the superpotential of the deformed gauge theory, however this is not immediately helpful from the matrix model point of view on gravitational corrections to the $\cN=2$ theories.

Another connection between matrix models and gauge theories or topological strings is provided by the formalism of Eynard and Orantin. In \cite{eyn-or} they showed how to solve the loop equations and derived a recursive procedure to get correlation functions and the partition function of a matrix model. In fact, this recursion and its solution is determined by the spectral curve and the associated differential alone. One can therefore start with \emph{any} curve and a differential and derive the corresponding set of symplectic invariants, without the knowledge of the matrix model, or even without the underlying matrix model. It was indeed checked in \cite{remodel} that such invariants, computed for a Seiberg-Witten curve or a B-model curve, agree with gauge theory and topological string invariants. However this relation to matrix models is not explicit, and the lack of the matrix model potential still prevents one from considering phenomena such as eigenvalue tunelling.

In this paper we take yet another approach and present how partition functions of a certain class of gauge theories and topological strings can be written as matrix models. A first step in this direction was taken in \cite{CGMPS}, where a matrix model for bundles over $\mathbb{P}^1$ was written down. This result was confirmed in \cite{eynard-planch}, where in more general context the Plancherel measure was rewritten as a matrix model. In \cite{SW-matrix} these results were generalized to pure $SU(n)$, $\cN=2$ theories in 4 and 5 dimensions. In 5-dimensional case it was also possible to take into account Chern-Simons terms, which gave rise to more complicated toric geometries and multi-matrix models. 

The matrix models which we derive here and those in \cite{SW-matrix} in principle encode \emph{all} gauge theory amplitudes $\cF_g$ for arbitrary $g$, without a need of taking any special limit or identification of parameters. The price for that is quite involved form of the matrix model potentials: while to get the correct $\cF_0$ and $\cF_1$ in the Dijkgraaf-Vafa case it is sufficient to choose a polynomial of order $(n+1)$ (to make eigenvalues condense in $n$ extrema and give rise to the genus $(n-1)$ Seiberg-Witten curve), the potentials which we derive abound in polylogarithms and their quantum generalizations. Of course these new potentials have the correct number of extrema to give rise to the curve of the appropriate genus (it was checked explicitly in \cite{SW-matrix} that the matrix model spectral curve agrees with the Seiberg-Witten curve for the pure $SU(2)$ theory).

In a wider context, in this paper we present a more general method of turning into matrix models certain expressions given by sums over partitions. This method is based on the (last line of the) formula (\ref{prod-hook}). We use this method in a particular case of $\cN=2^*$ theories.
The matrix models which we derive for such theories in various dimensions are given by formulae (\ref{DM-4dU1}), (\ref{DM-4dUn}), (\ref{DM-5dU1}), and (\ref{DM-6dU1}).

The plan of the paper is as follows. In section \ref{sec-hook-lenght} a construction of matrix models for a class of expressions involving discrete sums over partitions is presented. In section \ref{sec-4dim}  partition functions of 4-dimensional $2^*$ theories, obtained from instanton sums, are reformulated as matrix models. In section \ref{sec-5dim} a matrix model is derived for a partition function of the 5-dimensional theory obtained from the topological vertex calculus and related to the genus one curve. In section \ref{sec-6dim} we derive a matrix model for 6-dimensional theory, related to the genus two curve. Section \ref{sec-conclude} contains conclusions and discussion.


\section{The hook length formula and matrix models}  \label{sec-hook-lenght}

In this section we explain how to construct matrix models encoding discrete sums over partitions, with arbitrary dependence on the hook lengths. While the general idea of turning discrete sums into matrix integrals is the same as in \cite{eynard-planch,SW-matrix}, here we generalize it to a wide class of expressions. Such expressions arise in particular as partition functions of supersymmetric theories and topological strings. In the following sections we apply this method to the case of $\cN=2^*$ theories in various dimensions.

To start with we fix some notation. For a partition $R=(R_1,R_2,\ldots,R_N)$, with $R_1\geq R_2 \geq \ldots R_N \geq 0$, we define 
\be
h_i = R_i - i + N,    \label{hiRi}
\ee
which constitute a strictly decreasing sequence. The hook length of a box $\square=(i,j)\in R$ (with $i$ and $j$ denoting respectively the row and the column of a box) is defined as
\be
\hs=R_i+R_{j}^{t}-i-j+1, \label{hook}
\ee
where $R^t$ is a transposed matrix.

We now present a formula which is a basis of many calculations in what follows. 
For a function $\varphi$ such that $\varphi(n)\neq 0$ for $n\neq 0$ we have

\vspace{2 mm}

\noindent
\fbox{
\addtolength{\linewidth}{-2\fboxsep}%
\addtolength{\linewidth}{-2\fboxrule}%
\begin{minipage}{\linewidth}
\bea
\prod_{\square\in R}  \frac{1}{\varphi(h(\square))}  & = & \prod_{1\leq i<j <\infty} \frac{\varphi(R_i-R_j-i+j)}{\varphi(j-i)} = \prod_{1\leq i<j <\infty} \frac{\varphi(h_i-h_j)}{\varphi(j-i)} = \nonumber \\
& = & \Big(\prod_{1\leq i<j \leq N} \frac{\varphi(h_i-h_j)}{\varphi(j-i)} \Big) \Big(\prod_{j=1}^N \prod_{k=1}^{R_j} \frac{1}{\varphi(k-j+N)}   \Big) = \nonumber \\
& = & \Big(\prod_{1\leq i<j \leq N} \varphi(h_i-h_j) \Big) \Big(\prod_{j=1}^N \prod_{k=1}^{h_j} \frac{1}{\varphi(k)}   \Big). \label{prod-hook}
\eea
\end{minipage}
}

\vspace{2 mm}

The equalities in the first two lines were derived and used (among the others) in \cite{HIV,Zhou2003}. This is however the third line which immediately allows to relate expressions involving products over hook lengths with matrix models, and which we use repeatedly in next sections. To go from the second to the third line we notice that the first product in the denominator can be rewritten as $\prod_{1\leq i<j \leq N} \varphi(j-i) = \prod_{j=1}^{N-1}\big[\varphi(1)\cdots \varphi(j)\big]$. Then each expression in the square bracket combines with one of the products $\prod_{k=1}^{R_j} \varphi(k-j+N)$ into the product of the form $\prod_{k=1}^{h_j} \varphi(k)$. In total we get $N$ such expressions, each one for fixed $j = 1,\ldots,N$.

We are interested in deriving matrix models corresponding to the sums over partitions involving products of the above type. To find such models we first truncate sums over all partitions to sum over partitions of a fixed number of $N$ rows. Then we use the formula (\ref{prod-hook}) to replace the dependence on hook lengths by the dependence on $h_i$ defined in (\ref{hiRi}). The first factor in the third line $\prod \varphi(h_i-h_j)$ turns into some generalization of the Vandermonde determinant and becomes a part of the integration measure, while the product $\prod_{k=1}^{h_j} \frac{1}{\varphi(k)}$ gives rise to the matrix model potential $V$ (after writing it as an exponential). The product $\prod_{j=1}^N$ represents the trace.  We rewrite the above sum as continuous integrals by introducing an auxiliary function with poles in all integer values of $h_i$ and treating $u_i=\hbar h_i$ (or equivalently $u_i=-g_s h_i$ in five- and six-dimensional cases) as continuous matrix eigenvalus in the 't Hooft limit 
\be
\hbar\to 0, \qquad N\to\infty, \qquad \hbar N = const.    \label{tHooft}
\ee
Because of the rescaling of $h_i$, the dependence on $\hbar$ (or $g_s$) appears not only in front of the potential term in (\ref{mmodel}) but it also enters the potential in a nontrivial way. We therefore expand the potentials in a series in $\hbar$ (or $g_s$), in particular in order to identify the lowest order terms which survive the 't Hooft limit.

\bigskip

\underline{Example 1:} for $R=(6,3,1,1)$ one can check the formula (\ref{prod-hook}) explicitly, substituting in this case $(h_1,h_2,h_3,h_4) = (9,5,2,1)$ and the following hook lengths of all boxes:
$$
\young({{9}{6}{5}{3}{2}{1}},{{5}{2}{1}},{2},{1})
$$

\underline{Example 2:} for $\varphi(x)=x^2$ the product over hook lengths reproduces the Plancherel measure \cite{eynard-planch}. The equality (\ref{prod-hook}) implies then
\be
\sum_{R}\prod_{\square\in R} \frac{1}{\hs^2} = \frac{1}{N!} \sum_{h_1,\ldots,h_N} \prod_{1\leq i<j \leq N} (h_i-h_j)^2 \ \prod_{j=1}^N \frac{1}{h_j!^{\, 2}}.     \label{planch}
\ee
Introducing $u_i = \hbar h_i$ allows to write this as a (continuous) matrix model, with the Vandermonde determinant arising from the product
$\prod (h_i-h_j)^2$, and the potential given by the logarithm of the $\Gamma$ function (which arises as the analytic continuation of the factorial). This way we immediately recover the results of \cite{eynard-planch}.


\section{4-dimensional theories}    \label{sec-4dim}

In this section we derive a matrix model for 4-dimensional $2^*$ theory. This theory can be constructed from $\cN=4$ theory by adding mass $m$ to the hypermultiplet. This theory has a microscopic coupling $\tau_0$, and the associated parameter $Q=e^{2\pi i\tau_0}$ can be used in instanton counting. We also use the following notation:
\be
Q=e^{-T},\qquad  m = \hbar \mu,\qquad  u_i = \hbar h_i.     \label{rescale}
\ee
In general in what follows $h_i$ will be defined as in (\ref{hiRi}), while $u_i$ will become a continuous parameter after taking the limit (\ref{tHooft}).



First we consider $U(1)$ theory, or more precisely its twisted version, for which the partition function localizes on instanton configurations labeled by partitions $R$. It was shown in \cite{Nek-Ok} that this partition function can be written as a sum over partitions
\be
Z_{4d}^{U(1)} = \sum_R Q^{|R|} \prod_{\square\in R} \frac{\hs^2 - \mu^2}{\hs^2}.    \label{Z4dU1}
\ee
In this case each summand is a product over all boxes of $R$ and the dependence on each box is only via the hook length. Identifying $\varphi^{U(1)}_{4d}(x) = \frac{x^2}{x^2-\mu^2}$ we can now use our hook length formula. The first factor in (\ref{prod-hook}) becomes
$$
\prod_{1\leq i<j \leq N} \varphi^{U(1)}_{4d}(h_i-h_j) = \prod_{1\leq i<j \leq N} \frac{(h_i-h_j)^2}{(h_i-h_j-\mu)(h_i-h_j+\mu )} = \prod_{i\neq j}^N \frac{h_i-h_j}{h_i-h_j-\mu}.
$$

Using $\prod_{k=1}^h (k+\mu) = \frac{\Gamma(\mu + h + 1)}{\Gamma(\mu + 1)}$ we find the contribution to the potential
$$
\prod_{k=1}^h \frac{k^2-\mu^2}{k^2} = \frac{\Gamma(1+h+\mu)\, \Gamma(1+h-\mu)}{\Gamma(1+\mu)\, \Gamma(1-\mu)\, \Gamma(1+h)^2}.
$$

To replace sums by integrals over matrix eigenvalus, we introduce the function \cite{eynard-planch,SW-matrix}
$$
f(x) = -x \G(-x) \G(x) e^{-i \pi x} = \frac{\pi e^{-i\pi x}}{\sin(\pi x)},
$$
which has simple poles at all integer values of the argument. To find expansion of the potential in $\hbar$ we also make use of the expansion of the digamma function (i.e. the logarithmic derivative of the $\Gamma$ function) 
\be
\psi(x) = \frac{d}{dx} \log\,\G(x) = -\frac{1}{2x} + \log\, x - \sum_{m=1}^{\infty} \frac{B_{2m}}{2m x^{2m}},
\ee 
where $B_{2m}$ are Bernoulli numbers. 

Combining the above, and writing the second factor in equation (\ref{prod-hook}) as an exponent, we obtain the matrix model of the form (\ref{mmodel}) with 

\vspace{2 mm}

\noindent
\fbox{
\addtolength{\linewidth}{-2\fboxsep}%
\addtolength{\linewidth}{-2\fboxrule}%
\begin{minipage}{\linewidth}
\bea
\mathcal{D}M^{U(1)}_{4d} & = & \prod_{k=1}^N du_k \, \prod_{i\neq j}^N \frac{u_i-u_j}{u_i-u_j-m}  \label{DM-4dU1} \\
V^{U(1)}_{4d}(u) & = & T u + u \log\frac{u^2}{u^2-m^2} + m\log\frac{u-m}{u+m} + \nonumber \\
& & + \frac{\hbar}{2} \log\frac{u^2}{u^2-m^2} + \sum_{k=1}^{\infty} \frac{B_{2k} \hbar^{2k}}{2k(1-2k)}\Big(\frac{1}{(u+m)^{2k-1}} + \frac{1}{(u-m)^{2k-1}}  - \frac{2}{u^{2k-1}}  \Big) .\nonumber 
\eea
\end{minipage}
}

\vspace{2 mm}

\begin{figure}[htb]
\begin{center}
\includegraphics[width=0.5\textwidth]{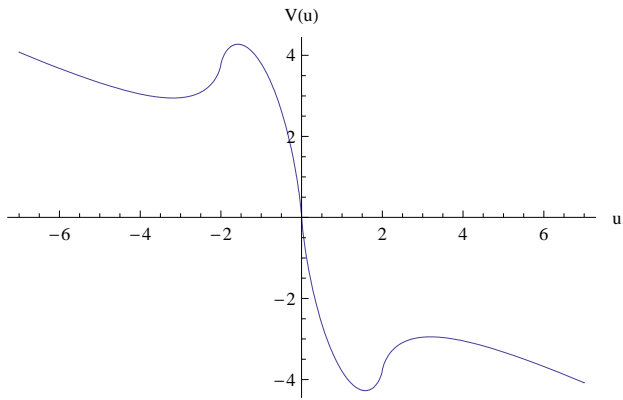} 
\begin{quote}
\caption{Planar limit of the matrix model potential for 4-dimensional, $U(1), 2^*$ theory.} \label{fig-V4dU1}
\end{quote}
\end{center}
\end{figure}

As expected the matrix model measure turns out to be a mass deformation of the Vandermonde determinant. The terms in the first line of the potential are the only terms essential in the planar limit and are shown in figure \ref{fig-V4dU1}. The terms in the last line are proportional to higher powers of $\hbar$ and vanish in the limit (\ref{tHooft}).

Now we consider a generalization of the above result to $U(n)$ theory. A particular point on the Coulomb branch we are interested in is specified by the values $a_l = \hbar p_l$ and we introduce $a_{lk}=a_l-a_k$. The instanton part of the partition function for $U(n)$ theory with the adjoint hypermultiplet reads \cite{Nek,Nek-Ok}
\bea
Z^{U(n)}_{4d} & = & \sum_{\vec{R}=(R^{(1)},\ldots,R^{(n)})} Q^{|\vec{R}|} Z_{\vec{R}},   \nonumber \\
Z_{\vec{R}} & = & \prod_{(l,i)\neq (k,j)} \frac{a_{lk}+\hbar(R^{(l)}_i - R^{(k)}_j + j - i)}{a_{lk}+\hbar(j-i)}
\frac{a_{lk}+m+\hbar(j-i)}{a_{lk}+m+\hbar(R^{(l)}_i - R^{(k)}_j + j - i)}. \label{Z4dUn}
\eea

This partition function involves a summation over $n$ partitions $R^{(k)}$ and explicitly is not of the form (\ref{prod-hook}). To deal with it we perform the same trick as in \cite{SW-matrix}. For each partition $R^{(k)}$ we introduce $h^{(k)}_i$ defined as a slight generalization of (\ref{hiRi}), 
$$
h^{(k)}_i  =  R_i-i+N+p_k,
$$
and then concatenate 
\be
h_{i=1,\ldots,nN} = (h^{(1)}_1,\ldots,h^{(1)}_N,h^{(2)}_1,\ldots,h^{(2)}_N,\ldots\ldots,h^{(n)}_1,\ldots,h^{(n)}_N).
\ee
This effectively leads to the one-matrix model with eigenvalues $u_i=\hbar h_i$. After manipulations similar to those in \cite{SW-matrix}, 
we obtain the matrix model of the form (\ref{mmodel}) with

\vspace{2 mm}

\noindent
\fbox{
\addtolength{\linewidth}{-2\fboxsep}%
\addtolength{\linewidth}{-2\fboxrule}%
\begin{minipage}{\linewidth}
\bea
\mathcal{D}M^{U(n)}_{4d} & = & \prod_{k=1}^{nN} du_k \, \prod_{i\neq j}^N \frac{u_i-u_j}{u_i-u_j-m}  \label{DM-4dUn} \\
V^{U(n)}_{4d}(u) & = & T u + \sum_{l=1}^n \Big[ (u-a_l)\log\frac{(u-a_l)^2}{(u-a_l)^2-m^2} + m\log\frac{u-a_l-m}{u-a_l+m} + \nonumber \\
& &  + \sum_{k=1}^{\infty} \frac{B_{2k} \hbar^{2k}}{2k(1-2k)}\Big(\frac{1}{(u-a_l+m)^{2k-1}} + \frac{1}{(u-a_l-m)^{2k-1}}  - \frac{2}{(u-a_l)^{2k-1}}  \Big) + \nonumber \\
& & + \frac{\hbar}{2} \log\frac{(u-a_l)^2}{(u-a_l)^2-m^2} \Big] . \nonumber
\eea
\end{minipage}
}

\vspace{2 mm}

\begin{figure}[htb]
\begin{center}
\includegraphics[width=0.5\textwidth]{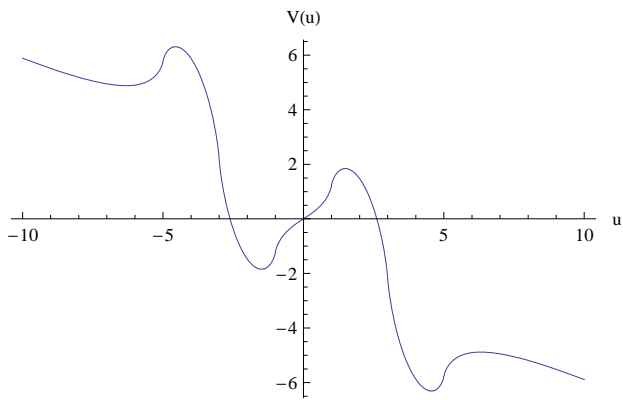} 
\begin{quote}
\caption{Planar limit of the matrix model potential for 4-dimensional, $U(3), 2^*$ theory.} \label{fig-V4dUn}
\end{quote}
\end{center}
\end{figure}

Again the measure is a mass deformation of the Vandermonde determinant, and only the terms in the first line of $V^{U(n)}_{4d}(u)$, shown in figure \ref{fig-V4dUn}, contribute in the planar limit.

In principle one could generalize the above 4-dimensional matrix model to 5 and 6 dimensions by replacing the instanton factor in (\ref{Z4dUn}) by its trigonometric and elliptic counterparts \cite{Nek-Ok,HIV}. This would lead to the corresponding generalizations of (\ref{DM-4dUn}). In what follows we will consider such generalizations only for $U(1)$ theory, using the results obtained from the topological vertex computations and stressing the relation of these theories to genus one and two curves.

\section{5-dimensional $U(1)$ theory}    \label{sec-5dim}

In this section we consider 5-dimensional $U(1)$ theory compactified on a circle of size $\beta$, taking the advantage of the topological vertex computations presented in \cite{HIV}. Following the topological string convention, we use the parameter $g_s$ instead of $\hbar$, as well as
$$
q=e^{-g_s},\qquad q^{h_i}=e^{-g_s h_i},\qquad  u_i=-g_s h_i,        
$$
with $u_i$ referring to the continuum matrix model counterpart of $h_i$ defined in (\ref{hiRi}). We also use the following notation for the deformed quantities
$$
[ h ] = q^{-h/2} -q^{h/2},  \qquad \qquad [ h ]!  =  \prod_{i=1}^h (q^{-i/2} - q^{i/2}).
$$

\begin{figure}[htb]
\begin{center}
\includegraphics[width=0.3\textwidth]{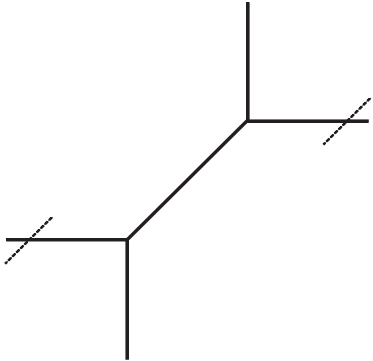}
\begin{quote}
\caption{Geometric realization of the 5-dimensional, $2^*, U(1)$ theory in terms of the periodic topological vertex diagram. Slashes on horizontal legs denote their identification along a circle.} \label{curve5d}
\end{quote}
\end{center}
\end{figure}

The instanton part of the partition function is related to the $(p,q)$ five-brane web or the topological vertex diagram shown in figure \ref{curve5d}. Two K{\"a}hler parameter of this diagram, $T$ and $T_m$, relate to the gauge coupling $\tau_0$ and the hypermultiplet mass $m$, so that
\be
\qquad Q_m = e^{-T_m}=e^{\beta m}, \qquad Q = e^{-T} = Q_m^{-1} e^{2\pi i \tau_0}.      \label{QQm}
\ee

The instanton partition function encoded in the diagram in figure \ref{curve5d} is given in terms of the topological vertex as
$$
Z^{U(1)}_{5d} = \sum_{P,R} (-Q)^{|R|} (-Q_m)^{|P|} C_{\bullet P^t R}(q) C_{P R^t \bullet} (q),
$$
with the amplitude $C_{PQR}$ derived in \cite{vertex}. This partition function was manipulated in \cite{HIV} to the form
\be
Z^{U(1)}_{5d} = \sum_R Q^{|R|} \prod_{\square\in R} \frac{(1-Q_m q^{\hs})(1-Q_mq^{-\hs})}{(1-q^{\hs})(1-q^{-\hs})}  \label{Z-U15d}
\ee
This expression is explicitly of the form (\ref{prod-hook}). If we define
\be
\varphi^{U(1)}_{5d}(x,Q) = \frac{(1-q^{x})(1-q^{-x})}{(1-Q q^{x})(1-Q q^{-x})},   \label{varphi5d}
\ee
a straightforward computation gives
\be
\prod_{1\leq i<j \leq N} \varphi^{U(1)}_{5d}(h_i-h_j,Q_m) = e^{-\frac{T_m N (N-1)}{2}} \prod_{i\neq j}^N \frac{\sinh\frac{u_i-u_j}{2}}{\sinh\frac{u_i-u_j-T_m}{2}}.
\ee
This gives rise to the measure of the matrix model we are after.



To derive a matrix model potential we use similar tools as in \cite{eynard-planch,SW-matrix}. 
We write the quantum dilogarithm as
$$
g(x) = \prod_{i=1}^{\infty} (1-x^{-1}q^i).
$$
It vanishes $g(q^h)=0$ for $h$ a positive integer, and at such points its derivative is
$$ 
g'(q^h) = - \frac{g(1)^2 e^{i\pi h} q^{-h(h-1)/2}}{q^h(1-q^h)g(q^{-h})}.
$$
Therefore the following function has simple poles with residue 1 for $x=q^h$ with $h\in\mathbb{N}$
$$
f_q(x) = -\frac{g(1)^2 e^{-\frac{i\pi}{g_s}\log x} e^{\frac{(\log x)^2}{2g_s}}}{(1-x)\sqrt{x}g(x)g(x^{-1})}.
$$
Let us also note that for $x=q^l$
\bea
[l]! & = & q^{-l(l+1)/4} \frac{g(1)}{g(x^{-1})}, \nonumber \\
\frac{1}{([l-p]!)^2} & = & \frac{g(x^{-1}q^p)^2}{g(1)^2} \exp\big(\frac{1}{2\log q} \log(xq^{-p}) 
\log(xq^{-p+1})\big). \nonumber 
\eea
Also note \cite{eynard-planch} that to get the expansion of the potential in $g_s$ one needs
$$
\log g(x) = -\frac{1}{g_s} \sum_{m=0}^{\infty} \Li_{2-m}(x^{-1}) \frac{B_{m}g_s^m}{m!},
$$
and in consequence, for $q^h\equiv e^u$, we have
\bea
\prod_{k=1}^h (1-Qq^k) = \frac{g\big(\frac{1}{Q}\big)}{g\big(\frac{1}{Qe^u}\big)}  = 
e^{-\frac{1}{g_s}\sum_{m=0}^{\infty} \frac{g_s^m B_m}{m!}\big(\Li_{2-m}(Q) - \Li_{2-m}(Q e^u) \big)}.
\eea

Combining the above, the potential for 5-dimensional theory arises from the following contribution to the formula (\ref{prod-hook})
\be
\prod_{k=1}^h \frac{1}{\varphi^{U(1)}_{5d}(k,Q_m)} = Q_m^h \frac{g(Q_m) g(Q_m^{-1})}{g(1)^2}
e^{-\frac{1}{g_s}\sum_{m=0}^{\infty} \frac{g_s^m B_m}{m!}\big(2\Li_{2-m}(e^u) - \Li_{2-m}(Q_m e^u) - \Li_{2-m}(Q_m^{-1} e^u) \big)}.
\ee



To conclude, rewriting the expression (\ref{Z-U15d}) using the above manipulations, and up to overall constants, we get the matrix model expression of the form (\ref{mmodel}), with $\hbar$ replaced by $g_s$, and with

\vspace{2 mm}

\noindent
\fbox{
\addtolength{\linewidth}{-2\fboxsep}%
\addtolength{\linewidth}{-2\fboxrule}%
\begin{minipage}{\linewidth}
\bea
\mathcal{D}M^{U(1)}_{5d} & = & \prod_{k=1}^N du_k \, \prod_{i\neq j}^N  \frac{\sinh\frac{u_i-u_j}{2}}{\sinh\frac{u_i-u_j-T_m}{2}}   \label{DM-5dU1}  \\
V^{U(1)}_{5d}(u) & = & -u(T+T_m) + 2\, \Li_2(e^u) - \Li_2(Q_me^u) - \Li_2(Q_m^{-1}e^u)  + \nonumber \\
& & + g_su + \frac{g_s}{2} \log\frac{(1-e^u)^2}{(1-Q_m e^u)(q-Q_m^{-1} e^u)} + \nonumber \\
& & + \sum_{k=1}^{\infty} \frac{B_{2k} g_s^{2k}}{(2k)!} \Big(2\, \Li_{2(1-k)}(e^u) - \Li_{2(1-k)}(Q_m e^u) - \Li_{2(1-k)}(Q_m^{-1} e^u)  \Big) .\nonumber
\eea
\end{minipage}
}

\vspace{2 mm}

\begin{figure}[htb]
\begin{center}
\includegraphics[width=0.5\textwidth]{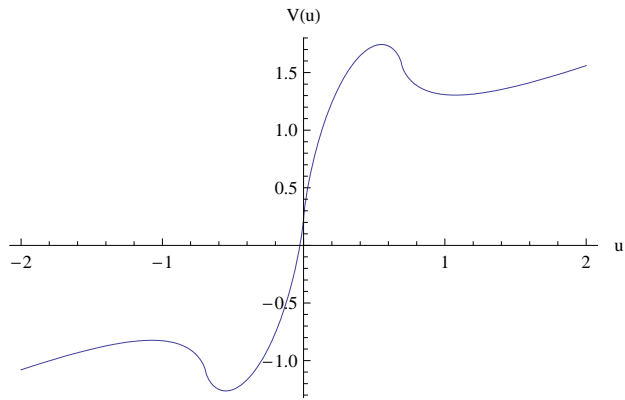}
\begin{quote}
\caption{Planar limit of the matrix model potential for 5-dimensional, $U(1), 2^*$ theory.} \label{fig-V5dU1}
\end{quote}
\end{center}
\end{figure}

As expected the measure of this matrix model is the trigonometric and massive deformation of the Vandermonde determinant. The first line of the potential $V^{U(1)}_{5d}(u)$ contains the only terms which contribute in the planar limit, which are shown in figure \ref{fig-V5dU1}.

\section{6-dimensional $U(1)$ theory}     \label{sec-6dim}

\begin{figure}[htb]
\begin{center}
\includegraphics[width=0.3\textwidth]{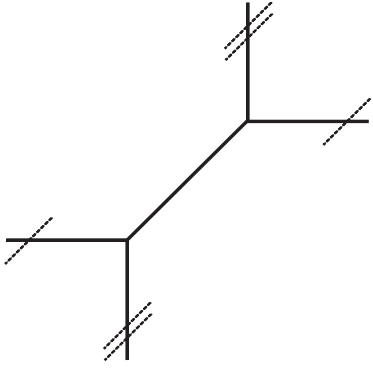}
\begin{quote}
\caption{Geometric realization of the 6-dimensional, $2^*, U(1)$ theory in terms of the doubly periodic topological vertex diagram. Slashes on horizontal and vertical legs denote their identification along a circle.} \label{curve6d}
\end{quote}
\end{center}
\end{figure}

Generalization of the $U(1), 2^*$ theory to 6 dimensions can be understood in terms of the toric diagram with two periodic directions, representing the curve of genus 2, as shown in figure \ref{curve6d}. In addition to parameters $Q$ and $Q_m$ introduced in (\ref{QQm}) and relevant for the situation in figure \ref{curve5d}, now there is an additional K{\"a}hler parameter 
$$
Q_1 = e^{-T_1}.
$$
The rest of the notation is the same as in section \ref{sec-5dim}.

Instanton partition function for 6-dimensional theory with $U(1)$ gauge group, related to the above genus two surface, is given by the topological vertex amplitude
$$
Z^{U(1)}_{6d} = \sum_{P,R,S} (-Q)^{|R|} (-Q_m)^{|P|} (-Q_1)^{|S|} C_{P R S}(q) C_{P^t R^t S^t} (q).
$$
This expression was manipulated in \cite{HIV} to the following form 
\bea
Z^{U(1)}_{6d} & = & \sum_R Q^{|R|} \prod_{\square\in R} \frac{(1-Q_m q^{\hs})(1-Q_mq^{-\hs})}{(1-q^{\hs})(1-q^{-\hs})} \times  \label{Z-U16d} \\
& \times & \prod_{k=1}^{\infty} \frac{(1-Q_{\rho}^k Q_m q^{\hs})(1-Q_{\rho}^k Q_m q^{-\hs})(1-Q_{\rho}^k Q_m^{-1} q^{\hs})(1-Q_{\rho}^k Q_m^{-1} q^{-\hs})}{(1-Q_{\rho}^k q^{\hs})^2(1-Q_{\rho}^k q^{-\hs})^2}.     \nonumber
\eea
where
$$
Q_{\rho}=Q_1 Q_m.
$$
This again involves summands of the form (\ref{prod-hook}) which allows to make contact with matrix models.
We define
\be
\varphi_{6d}^{U(1)}(x) = \varphi_{5d}^{U(1)}(x,Q_m) \prod_{k=1}^{\infty} \frac{(1-Q_{\rho}^k q^{x})^2(1-Q_{\rho}^k q^{-x})^2}{(1-Q_{\rho}^k Q_m q^{x})(1-Q_{\rho}^k Q_m q^{-x})(1-Q_{\rho}^k Q_m^{-1} q^{x})(1-Q_{\rho}^k Q_m^{-1} q^{-x})},
\ee
with $\varphi_{5d}^{U(1)}(x,Q_m)$ given in (\ref{varphi5d}).

The measure of the 6-dimensional matrix model arises from the familiar by now products over $\varphi_{6d}^{U(1)}(h_i - h_j)$, which after some manipulations become
\be
\prod_{1\leq i<j \leq N} \varphi_{6d}^{U(1)}(h_i - h_j) = e^{-T_m N(N-1)}\prod_{i\neq j}^N \frac{\theta_1(q^{h_i-h_j} | Q_{\rho})}{\theta_1(Q_m q^{h_i-h_j} | Q_{\rho})}
\ee
with the notation
$$
\theta_1(t|Q) = i\sum_{n\in\mathbb{Z}} (-1)^n Q^{\frac{n(n+1)}{2} + \frac{1}{8}} t^{n+\hf}.
$$
The manipulations necessary to get the above measure include in particular replacing infinite products over $k$ in $\varphi_{6d}^{U(1)}$ by the theta function using the following form of the Jacobi triple product identity
$$
\prod_{k=1}^{\infty} (1-Q^k t)(1-Q^{k} t^{-1}) (1-Q^k) = \frac{\theta_1(t|Q)}{i Q^{1/8} (t^{1/2}-t^{-1/2})} \, .
$$

The matrix model potential arises from the terms of the form
$$
\prod_{l=1}^{\infty} \prod_{k=1}^h \frac{1-Q^l q^k}{1-Q^l q^{-k}} 
= \Big[\prod_{l=1}^{\infty} (-Q^l)^h q^{-\frac{h(h+1)}{2}} \frac{g(Q^{-l})}{g(Q^l)} \Big]
 e^{-\frac{1}{g_s} \sum_{k=0}^{\infty} \frac{g_s^k B_k}{k!}\mathbb{L}_{3-k}(q^h,Q)},
$$
where we introduced the quantum polylogarithm 
\be
\mathbb{L}_{N+1}(x,Q) = \sum_{n\in\mathbb{Z}_{\neq 0}} \frac{x^{|n|}}{(1-Q^n)|n|^N}.
\ee
Special cases of this function include the quantum dilogarithm $\mathbb{L}_2$ considered in \cite{dilog}, as well as  the quantum trilogarithm $\mathbb{L}_3$ which appears in \cite{trilog1,trilog2,Nek-0901}.

Collecting the above results, and after some more algebraic manipulations, we obtain the 6-dimensional matrix model

\vspace{2 mm}

\noindent
\fbox{
\addtolength{\linewidth}{-2\fboxsep}%
\addtolength{\linewidth}{-2\fboxrule}%
\begin{minipage}{\linewidth}
\bea
\mathcal{D}M_{6d}^{U(1)} & = & \prod_{k=1}^N du_k \, \prod_{i\neq j}^N  \frac{\theta_1(e^{u_i-u_j} | Q_{\rho})}{\theta_1(e^{u_i-u_j-T_m} | Q_{\rho})}      \label{DM-6dU1}  \\
V_{6d}^{U(1)}(u) & = & -u(T + T_m) + g_s u + \sum_{k=0}^{\infty} \frac{B_{k} g_s^{k}}{(k)!} \Big(2\, \Li_{2-k}(e^u) - \Li_{2-k}(Q_m e^u) - \Li_{2-k}(Q_m^{-1} e^u)  \nonumber \\
& & -2\, \mathbb{L}_{3-k}(e^u,Q_{\rho}) + \mathbb{L}_{3-k}(Q_m e^u,Q_{\rho}) + \mathbb{L}_{3-k}(Q_m^{-1} e^u,Q_{\rho}) \Big). \nonumber
\eea
\end{minipage}
}

\vspace{2 mm}

The measure of this potential is the expected elliptic and massive generalization characteristic for compactified 6-dimensional theories \cite{HIV,Nek-0901}. In the planar limit only the terms in $V_{6d}^{U(1)}(u)$ proportional to $g_s^0\sim 1$ contribute, i.e. only the linear term and the first term for $k=0$ from the infinite sum, which involves the dilogarithms $\Li_2$ and quantum trilogarithms $\mathbb{L}_3$.


\section{Discussion} \label{sec-conclude}

In this paper we derived matrix models for $2^*$ theories in 4, 5 and 6 dimensions starting from expressions for partition functions obtained from instanton counting or topological vertex calculations. The measure of these matrix models is respectively of rational, trigonometric and elliptic type and includes also the massive deformation. The potentials are quite involved, and their expansion in $\hbar$ (or $g_s$) involves ordinary and quantum polylogarithms.

Our matrix models can also be regarded as complementing the results of \cite{HIV}. In that paper matrix models, topological vertex calculations and elliptic genera were related to each other, also in the context of $2^*$ theories. The matrix models discussed there are however of Dijkgraaf-Vafa type and are obtained starting from $1^*$ theory, i.e. $\mathcal{N}=1$ theory with three adjoint chiral superfields $\Phi_{i=1,2,3}$ of mass $m$ and the tree-level superpotential $\Tr(\Phi_1[\Phi_2,\Phi_3])$. As explained in the introduction in this case there are difficulties in relating higher genus matrix model $F_g$ to gauge theory amplitudes $\cF_g$. In contrary, by the very construction, the matrix models for $2^*$ derived here explicitly encode all gauge theory $\cF_g$.

It is natural to expect that the method we used can still be generalized. An interesting feature of the matrix models presented here is the fact that the formula (\ref{prod-hook}) implies that both the measure and the potential of the matrix model are encoded and derived from the same function $\varphi$. To obtain matrix models in yet more general situations would require relaxing this feature. One aim of this program would be to convert into matrix models topological string partition functions for local Calabi-Yau manifolds encoded in arbitrary toric diagrams. 

On the other hand, the use of the formula (\ref{prod-hook}) is not limited to the gauge theory or topological string partition functions, and with its help matrix models could be derived for other interesting physical and mathematical systems. Some such applications related to the simplest case of the Plancherel measure in the formula (\ref{planch}) were discussed in \cite{eynard-planch}. As for the next-to-simplest case of the 4-dimensional $2^*$ theory (\ref{Z4dU1}) let us just note, that some interest aroused in the field of enumerative combinatorics \cite{Han-NO,Stanley-NO} due to the formula found by Nekrasov and Okounkov in \cite{Nek-Ok}
$$
Z^{U(1)}_{4d} = \prod_{n=1}^{\infty} (1-Q^n)^{\mu^2 -1}.
$$
It would be interesting to see how combinatorial features of this formula relate to the combinatorics of diagrams of the matrix model (\ref{DM-4dU1}).

Our results should also be of interest from several other points of view. On one hand, Seiberg-Witten theories with matter, including $SU(2), 2^*$ theory in 4 dimensions, were considered recently in \cite{SWmatter} from the viewpoint of holomorphicity and modularity. Also the relation to the formalism of Eynard-Orantin was discussed there. Relating matrix models derived here to those results would be an interesting check of all these approaches.

On the other hand, $\cN=2$ theories in various dimensions are related to integrable systems. In 4, 5 and 6 dimensions theories these are respectively Calogero-Moser system \cite{DW}, Ruijsenaars-Schneider system \cite{Nek-integrable,BMMM-Ruij}, and their generalization considered in \cite{HIV,bra-hol}. Recently the new viewpoint on the relation to integrable systems was given in \cite{Nek-0901}. It would be of interest to understand relations between matrix models derived here and those integrable systems. 

$2^*$ theories with $U(1)$ gauge group in 5 and 6 dimensions, which correspond to toric geometries representing curves of genus one and two, were also considered recently in the context of the refined vertex \cite{refined-vertex} and related to cylindric partitions \cite{refined-U1,refined-U1branes}. One might wonder if there is any way of obtaining such a refined information from more general matrix models.

Finally, the genus two curve relevant for 6-dimensional $U(1)$ theory arises also in the context of dyon counting in $\cN=4$ string theory \cite{DVV-dyons,rerecounting}. The dyon counting function is given by the Igusa cusp form of weight 10, and in \cite{dhsv} it was shown to arise also as the genus one contribution to the partition function for the genus two curve. Finding any relations between both these systems, as well as between matrix models derived here and $\mathcal{D}$-modules analyzed in \cite{dhsv,dhs}, is certainly worth more effort.


\bigskip

\bigskip

\centerline{\Large{\bf Acknowledgments}}

\bigskip

I would like to thank Mina Aganagic, Rainald Flume, Hartmut Monien and Albrecht Klemm for inspiring discussions. I am also grateful to Rainald Flume and Albrecht Klemm for careful reading of the manuscript and many valuable comments. I thank University of California, Berkeley and University of Southern California for kind hospitality. This research was supported by the Humboldt Fellowship.


\bigskip 



\begin{thebibliography}{99}

\bibitem{DV}
R. Dijkgraaf, C. Vafa,
\emph{A Perturbative Window into Non-Perturbative Physics},
\textsf{hep-th/0208048}.

\bibitem{cachazo-vafa}
F. Cachazo, C. Vafa,
\emph{N=1 and N=2 Geometry from Fluxes},
\textsf{hep-th/0206017}.

\bibitem{gaugedmm}
R. Dijkgraaf, S. Gukov, V. Kazakov, C. Vafa,
\emph{Perturbative Analysis of Gauged Matrix Models},
\textsf{Phys. Rev. D68 (2003) 045007 [hep-th/0210238]}.

\bibitem{kmt}
A. Klemm, M. Marino, S. Theisen,
\emph{Gravitational corrections in supersymmetric gauge theory and matrix models},
\textsf{JHEP 0303 (2003) 051 [hep-th/0211216]}.

\bibitem{DST}
R. Dijkgraaf, A. Sinkovics, M. Temurhan,
\emph{Matrix Models and Gravitational Corrections},
\textsf{Adv. Theor. Math. Phys. 7 (2004) 1155-1176 [hep-th/0211241]}.

\bibitem{SW-1}
N. Seiberg, E. Witten,
\emph{Monopole Condensation, And Confinement In N=2 Supersymmetric Yang-Mills Theory},
\textsf{Nucl. Phys. B426 (1994) 19-52 [hep-th/9407087]}.

\bibitem{SW-2}
N. Seiberg, E. Witten,
\emph{Monopoles, Duality and Chiral Symmetry Breaking in N=2 Supersymmetric QCD},
\textsf{Nucl. Phys. B431 (1994) 484-550 [hep-th/9408099]}.

\bibitem{Nek}
N. Nekrasov,
\emph{Seiberg-Witten prepotential from instanton counting},
\textsf{Adv. Theor. Math. Phys. 7 (2004) 831-864 [hep-th/0206161]}.

\bibitem{Nek-Ok}
N. Nekrasov, A. Okounkov,
\emph{Seiberg-Witten theory and random partitions},
\textsf{hep-th/0306238}.

\bibitem{geom-eng}
S. Katz, A. Klemm, C. Vafa,
\emph{Geometric engineering of quantum field theories},
Nucl. Phys. {\bf B497} (1997) 173-195  [\textsf{hep-th/9609239}].

\bibitem{suN-Amir}
A. Iqbal, A. Kashani-Poor,
\emph{SU(N) Geometries and Topological String Amplitudes},
\textsf{Adv. Theor. Math. Phys. 10 (2006) 1-32 [hep-th/0306032]}.

\bibitem{HIV}
T. Hollowood, A. Iqbal, C. Vafa, 
\emph{Matrix Models, Geometric Engineering and Elliptic Genera},
\textsf{JHEP 0803 (2008) 069 [hep-th/0310272]}.

\bibitem{Zhou2003}
J. Zhou,
\emph{Curve counting and instanton counting},
\textsf{math.AG/0311237}.

\bibitem{Zhou2004}
J. Li, K. Liu, J. Zhou,
\emph{Topological String Partition Functions as Equivariant Indices},
\textsf{math.AG/0412089}.

\bibitem{vertex}
M. Aganagic, A. Klemm, M. Marino, C. Vafa
\emph{The Topological Vertex},
\textsf{Commun. Math. Phys. 254 (2005) 425-478 [hep-th/0305132]}.

\bibitem{adkmv}
M.~Aganagic, R.~Dijkgraaf, A.~Klemm, M.~Marino, C.~Vafa,
\emph{Topological strings and integrable hierarchies},
\textsf{hep-th/0312085}.

\bibitem{kmr}
A. Klemm, M. Marino, M. Rauch, \emph{in preparation}.

\bibitem{c-deform}
H. Ooguri, C. Vafa,
\emph{The C-Deformation of Gluino and Non-planar Diagrams},
\textsf{Adv. Theor. Math. Phys. 7 (2003) 53-85 [hep-th/0302109]}.

\bibitem{eyn-or}
  B.~Eynard, N.~Orantin,
  \emph{Invariants of algebraic curves and topological expansion},
  \textsf{math-ph/0702045}.

\bibitem{remodel}
V. Bouchard, A. Klemm, M. Marino, S. Pasquetti,
\emph{Remodeling the B-model},
\textsf{Commun. Math. Phys. 287 (2009) 117-178 [0709.1453 [hep-th]]}.

\bibitem{CGMPS}
N. Caporaso, L. Griguolo, M. Marino, S. Pasquetti, D. Seminara,
\emph{Phase transitions, double-scaling limit, and topological strings},
\textsf{Phys. Rev. D75 (2007) 046004 [hep-th/0606120]}.

\bibitem{eynard-planch}
B. Eynard,
\emph{All order asymptotic expansion of large partitions},
\textsf{0804.0381 [math-ph]}.

\bibitem{SW-matrix}
A. Klemm and P. Su{\l}kowski,
\emph{Seiberg-Witten theory and matrix models},
\textsf{0810.4944 [hep-th]}.

\bibitem{dilog}
L. Faddeev, R. Kashaev,
\emph{Quantum Dilogarithm},
\textsf{Mod.Phys.Lett. A9 (1994) 427-434 [hep-th/9310070]}.

\bibitem{trilog1}
A. Beilinson, A. Levin,
\emph{Elliptic Polylogarithms},
\textsf{Proceedings of Symposia in Pure Mathematics,
Vol.55, Part 2 (1994) 126-196}.

\bibitem{trilog2}
V. Kuznetsov, F. Nijhoff, E. Sklyanin, 
\emph{Separation of variables for the Ruijsenaars system},
\textsf{Commun. Math. Phys. 189 (1997) 855-877 [solv-int/9701004]}.

\bibitem{Nek-0901}
N. Nekrasov, S. Shatashvili,
\emph{Supersymmetric vacua and Bethe ansatz},
\textsf{hep-th/0901.4744}.

\bibitem{Han-NO}
G. Han,
\emph{The Nekrasov-Okounkov hook length formula: refinement, elementary proof, extension and applications},
\textsf{0805.1398 [math.CO]}.

\bibitem{Stanley-NO}
R. Stanley,
\emph{Some Combinatorial Properties of Hook Lengths, Contents, and Parts of Partitions},
\textsf{0807.0383 [math.CO]}.

\bibitem{SWmatter}
M.x. Huang, A. Klemm,
\emph{Holomorphicity and Modularity in Seiberg-Witten Theories with Matter},
\textsf{0902.1325 [hep-th]}.

\bibitem{DW}
R. Donagi, E. Witten,
\emph{Supersymmetric Yang-Mills Systems And Integrable Systems},
\textsf{Nucl. Phys. B460 (1996) 299-334 [hep-th/9510101]}.

\bibitem{Nek-integrable}
N. Nekrasov,
\emph{Five Dimensional Gauge Theories and Relativistic Integrable Systems},
\textsf{Nucl. Phys. B531 (1998) 323-344 [hep-th/9609219]}.


\bibitem{BMMM-Ruij}
H. Braden, A. Marshakov, A. Mironov, A. Morozov, 
\emph{The Ruijsenaars-Schneider model in the Context of Seiberg-Witten theory},
\textsf{Nucl. Phys. B558 (1999) 371-390 [hep-th/9902205]}.

\bibitem{bra-hol}
H. Braden, T. Hollowood,
\emph{The Curve of Compactified 6D Gauge Theories and Integrable Systems},
\textsf{hep-th/0311024}.

\bibitem{refined-vertex}
A. Iqbal, C. Kozcaz, C. Vafa,
\emph{The Refined Topological Vertex},
\textsf{hep-th/0701156}.

\bibitem{refined-U1}
A. Iqbal, C. Kozcaz, K. Shabbir,
\emph{Refined Topological Vertex, Cylindric Partitions and the U(1) Adjoint Theory},
\textsf{0803.2260 [hep-th]}.

\bibitem{refined-U1branes}
A. Iqbal, C. Kozcaz, T. Sohail,
\emph{Cylindric Partitions and Branes},
\textsf{0903.0961 [hep-th]}.

\bibitem{DVV-dyons}
R. Dijkgraaf, E. Verlinde, H. Verlinde,
\emph{Counting Dyons in N=4 String Theory},
\textsf{Nucl. Phys. B484 (1997) 543-561 [hep-th/9607026]}.

\bibitem{rerecounting}
D. Gaiotto,
\emph{Re-Recounting Dyons in N=4 String Theory},
\textsf{hep-th/0506249}.

\bibitem{dhsv}
R.~Dijkgraaf, L.~Hollands, P. Su{\l}kowski, C.~Vafa,
\emph{Supersymmetric gauge theories, intersecting branes and free fermions},
\textsf{JHEP 0802 (2008) 106  [0709.4446 [hep-th]]}.

\bibitem{dhs}
R.~Dijkgraaf, L.~Hollands, P. Su{\l}kowski,
\emph{Quantum Curves and $\mathcal{D}$-modules},
\textsf{0810.4157 [hep-th]}.


\end{thebibliography}
\end{document}